\documentstyle[12pt,epsfig]{article}

\def\numunux{\nu_\mu\rightarrow \nu_x}

 \newlength{\dinwidth}
\newlength{\dinmargin}
\def\Journal#1#2#3#4{{#1} {\bf #2}, #3 (#4)}
\def\etal{{\it et\ al.}}

\def\NIMA{{\em Nucl. Instrum. Methods} A}

\def\PRD{{\em Phys. Rev.} D}

\begin{document}

\title{Measurement of Atmospheric Neutrino\\
Oscillations with a High-Density Detector}
\maketitle

\vskip0.5cm

\begin{center}
A.~Curioni$^1$, G.~Mannocchi$^{2,3}$, L.~Periale$^2$, 
P.~Picchi$^{4,2,3}$, F.~Pietropaolo$^5$, S.~Ragazzi$^1$
\end{center}
\vskip1.0cm
\noindent
\hspace*{-0.127cm}$~^1$ Dipartimento di Fisica dell' Universit\`a and INFN, 
via Celoria 16, Milano, Italy\\
$~^2$  Istituto di Cosmogeofisica del C.N.R., 
corso Fiume 4, Torino, Italy\\
$~^3$ INFN, Laboratori Nazionali di Frascati, 
via E. Fermi 40, Frascati (Roma), Italy\\
$~^4$ Dipartimento di Fisica dell' Universit\`a,
via Giuria 1, Torino, Italy\\
$~^5$ INFN, Sezione di Padova, via Marzolo 8, Padova, Italy\\
\vspace{0.7 cm}

\abstract{We propose an experiment to test the hypothesis that
the reported anomaly on atmospheric neutrino fluxes is due to
$\numunux$ oscillations.
It will rely both on a disappearance technique, exploiting the
method of the dependence of the event rate on
$L/E$, which was recently shown to be effective
for detection of neutrino oscillation and measurement of
the oscillation parameters, and on an appearance technique,
looking for an excess of muon-less events at high energy produced
by upward-going tau neutrinos.
The detector will consist of iron
planes interleaved by limited streamer tubes. The
total mass will be about 30 $kt$. The possibility  of
recuperating most of the instrumentation from existing detectors
allows to avoid R\&D phases and to reduce construction time.
In four years of data taking, this experiment will be
sensitive to oscillations $\numunux$ with
$\Delta m^2 > 10^{-4}\ eV^2$ and a mixing
near to maximal, and answer the question whether
$\nu_x$ is a sterile or a tau neutrino.}


\newpage

\section{Introduction}

Recent Super-Kamiokande data \cite{superkamiokande}
confirm the existence of an anomaly
in the atmospheric muon neutrino fluxes, which is best
interpreted as a $\numunux$ oscillation, with a mixing near to
maximal and $\Delta m^2$ in the range $10^{-3}-10^{-2}\ eV^2$.
The non observation of a corresponding anomaly in the
electron neutrino fluxes and data from reactor experiments,
indicate that the oscillation either concern the muon and tau
neutrino or the muon and a new sterile neutrino.
This result, given its relevance, should be tested by an
independent experiment.
We believe that an experiment planned with this goal should have,
in itself, enough redundancy to be able to prove, or disprove,
that an observed anomaly in atmospheric neutrino fluxes be due to
neutrino oscillations.

It has recently been shown \cite{PioFrancesco} that the study of the 
atmospheric $\nu_\mu$ event rate as a function of the ratio
$L/E$, between the neutrino path length and its energy,
is an effective method for the detection of an atmospheric
$\nu_\mu$ deficit, if the deficit is due to oscillations.
Moreover this method
provides a clean measurement of the  oscillation parameters if
the mixing angle is large and
$10^{-4} < \Delta m^2 < 5\times 10^{-3}\ eV^2$.
In order to apply this method one should be able to measure, on
an event by event basis, both the energy $E$ and the direction of
the neutrino, from which the flight length $L$ is obtained.

For higher values of $\Delta m^2$ the modulation becomes too
fast to be detected by this method, but $\nu_{\mu}$ oscillations
would result in a detectable deficit of upward muon events
with respect to the downward ones.


In this paper we shall also discuss a different and independent method
which can be used to detect an appearance of muon-less
events produced by upward-going $\nu_\tau$, and we shall show how it can
distinguish between oscillation to tau or sterile neutrino.

\section{Choice of the Detector}

A water cherenkov is best suited for
events with simple topologies, in the quasi-elastics and
resonances region,
and has a high efficiency for electron identification.
A detector suitable for the application of the
methods outlined above should instead
be efficient on high energy events,
in the region of deep inelastic scattering,
and have a high capability of distinguishing muons from pions.
Moreover, having no interest in the study of oscillations
involving electron neutrinos, a detector filtering the
electro-magnetic component can be effectively chosen.
These considerations favour the choice of a high-density
tracking calorimeter.

In a tracking calorimeter the muon
energy can be measured either by means of a
magnetic field (magnetised iron), or from the range for muons
stopping in the detector. We believe that the use of a magnetic
field becomes too complex in an apparatus of the size needed for
atmospheric neutrinos, since the requirements of a magnetic field
over a large volume and of several high precision points
along the track, raise the costs and conflicts with high density.
The measurement of muon energy by range is instead free of
intrinsic difficulties, apart the requirement that the muon stops
inside the detector, which becomes critical for
detectors of low density.


There remains to be solved the problem of the identification of
the direction of flight of the incoming neutrino. In this respect
a tracking calorimeter is in general weaker than a cherenkov.
In the specific case of the events in which we are interested,
the problem can be solved as follows. In interactions producing
muons with energy above $\simeq 1\ GeV$ stopping in
the detector, the direction of the incoming neutrino can be
identified either by means of a visible event vertex, marked by
hadronic activity at one extreme of the muon track,
or, when no vertex is clearly visible, by increasing
residuals, due to multiple scattering, in the muon track fit.
For high-energy muon-less events
the direction, upward or downward, of the incoming neutrino
is recognised when the topology of the event allows to
identify the vertex.
The efficiency of this method increases for finer samplings.
A fine sampling, however, raises the cost and reduce the
detector density, thus limiting
the detector mass and reducing the stopping
power for muons.
We found that an overall optimisation
of event rates and efficiencies is obtained for
a relatively coarse sampling.


In conclusion a detector, suitable for the method of
analysis that we propose, should be designed according to the
following guidelines:
\begin{itemize}
\item total mass around 30 $kt$; 
\item structure of a high-density tracking calorimeter, capable
of measuring muon energy up to several GeV from its range;
\item precision of $\simeq 1\ mm$ on points along an isolated 
track, for the measurement of multiple scattering.
\end{itemize}

Moreover we would like to keep within reasonable limits the cost
and the construction time; therefore we believe that a well
tested technique requiring no R\&D should be adopted.

\section{The Detector Structure}
The detector will consist of 3 identical super-modules of
$12\times 12\times 13.2\ m^3$ each. 
Each super-module will consist of a stack of 120 iron planes
8 cm thick
interleaved by planes of limited streamer tubes (LST). The LST
will be of the same size as those used
in MACRO~\cite{MACRO}, that is $12\ m$ long and 
$3\times 3\ cm^2$ cross section.
They will provide two coordinates by means of the read out of anodes
and of diagonal strips, as in MACRO. The read-out and
acquisition chain will be the same as adopted by MACRO. Moreover
the anode signals will also be sent to TDC's with $15\ ns$
resolution, thus providing a resolution on the
x-coordinate better than $1\ mm$. 
Since each neutrino event will hit only a
limited part of the detector, several anode outputs
will be latched to a single TDC channel, thus limiting the number
of TDC channels.
The total area of LST planes will be $51840\ m^2$, about half of
which could be recuperated from MACRO.
The total mass of the detector will exceed $30\ kt$.

\section{Detection of Atmospheric Neutrino Oscillations}
Detection of oscillation of atmospheric neutrinos and
measurement of their parameters will rely on two main
techniques:
\begin{itemize}
\item disappearance of events with a high-energy muon pointing upward;
\item comparison of rates upward and downward muon-less
events of high energy.
\end{itemize}

%
%

We point out that while the first technique will test the
hypothesis of $\nu_{\mu}$ oscillations, and measure $\Delta m^2$
provided it is smaller than $5\times 10^{-3}\ eV^2$, the
second one will be used to discriminate between oscillations
to a sterile or a tau neutrino.
 
\subsection{Disappearance of high-energy muons}

It has been shown \cite{PioFrancesco,98-004} that
an effective method to test whether the anomaly
in the atmospheric neutrino flux reported by Super-Kamiokande
\cite{superkamiokande} is due to $\nu_\mu$ oscillation,
consists in searching for a modulation in the $\nu_\mu$ rate
plotted versus $L/E$. The modulation would be produced by
a disappearance probability given by:
\begin{equation}\label{prob}
P(\nu_\mu\rightarrow\nu_x)=\sin^2(2\Theta)\sin^2(1.27 \Delta m^2 L/E)
\end{equation}
the modulation period would thus be inversely proportional to 
$\Delta m^2$.
It has also been shown that by this method a measurement of
$\Delta m^2$ in the range between
$10^{-4}$ and $5\times 10^{-3}\ eV^2$ is affordable.
Moreover the method has the advantage of being
practically insensitive to the precise knowledge of 
the atmospheric neutrino flux, since the oscillation pattern is found by 
dips in the $L/E$ distribution,
while the atmospheric neutrino interaction
spectrum is known to be a 
slowly varying function of $L/E$.
The experimental requirement is that $L/E$ is measured with an error
much smaller than the modulation period, which decreases for increasing
values of $\Delta m^2$. This translates into requirements on
energy and angular resolution of the detector, which become more stringent
for higher $\Delta m^2$ values.

The detector that we propose has a hadronic energy resolution of
$\sigma(E)/E = 150\%/\sqrt(E)$, and essentially no capability
of reconstructing the hadron direction.

In the experiment simulation we reject all the events that are not
fully contained in the detector. The muon energy, obtained 
from range with errors due to
straggling and to the uncertainty on range measurement,
and the muon direction, obtained by a straight line fit to the
first meter of the muon track,
are measured with high precision. 

The condition of a good precision
on $L/E$ thus translates in the request that the muon direction
and the total reconstructed energy reproduce, within the quoted errors,
the neutrino direction and energy.

The request on the muon direction is simply satisfied
by the selection of events in which the hadronic energy is only a small
fraction of the total deposited energy. This request
can change with increasing neutrino energy; in fact the Lorentz boost
in the interaction is such that in high energy events the muon keeps
the original neutrino direction even if its fractional energy is low.
We have implemented this requirement imposing a cut on the hadronic 
fractional energy proportional to the total measured energy.

\begin{figure}[tb]
\begin{center}
\mbox{\epsfig{file=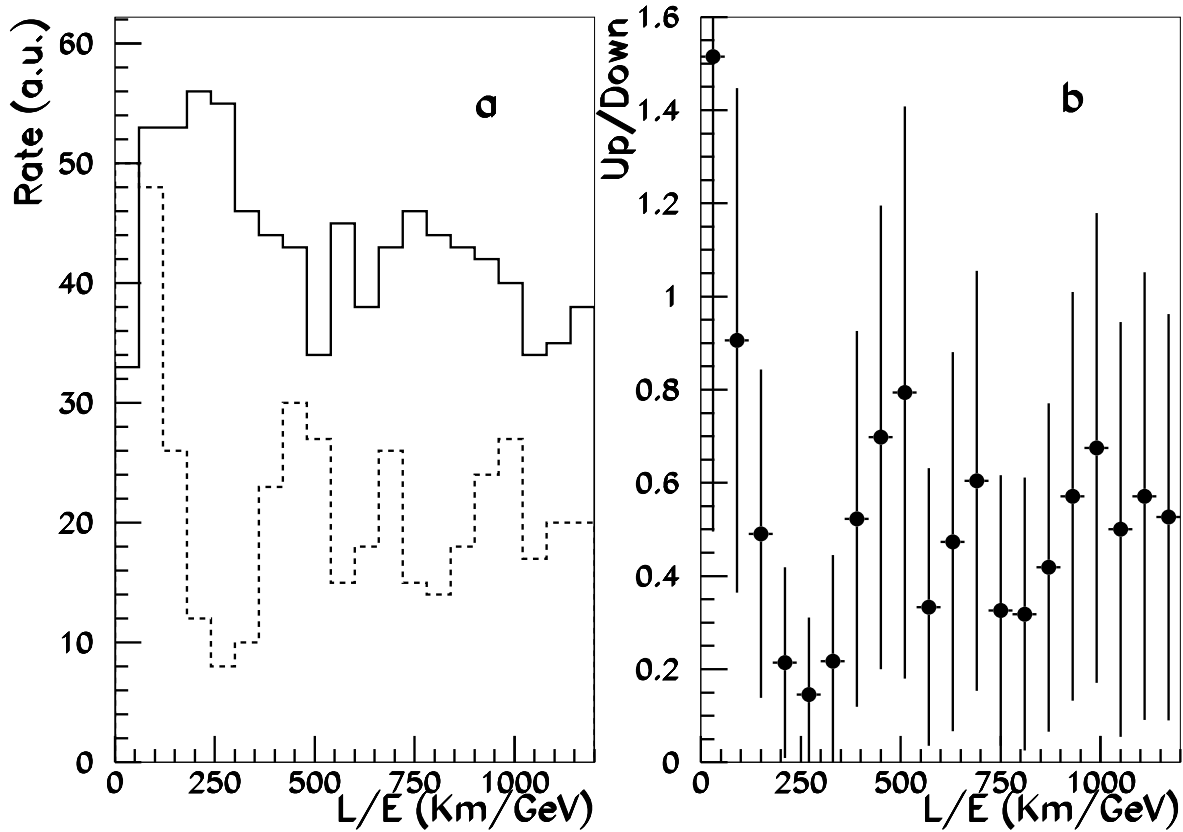,width=\linewidth}}
\end{center}
\caption{Measured $L/E$ distribution in presence
of $\nu_\mu\rightarrow\nu_x$ oscillations, with parameters
$\Delta m^2 = 5\times 10^{-3}\ eV^2$ and $\sin^2 (2\Theta)=1.0$
for upward muon events (dashed line) and downward ones (continuous line)
(a) and their ratio $R$ (b).
Events have been generated with high statistics,
error bars corresponding to the statistical uncertainty after
4 years of running are shown.
In order to compare the upward going neutrino sample 
with the downward going one as a function of $L/E$, we have assigned
to the downward going neutrinos (zenith angle $\theta < \pi / 2$) 
the distance they would have traveled if $\theta = \pi - \theta$. 
The ratio of the two distributions is thus 
approximately flux independent and the oscillating pattern is then 
put in evidence.
}
\label{fig:one}
\end{figure}

\begin{figure}[tb]
\begin{center}
\mbox{\epsfig{file=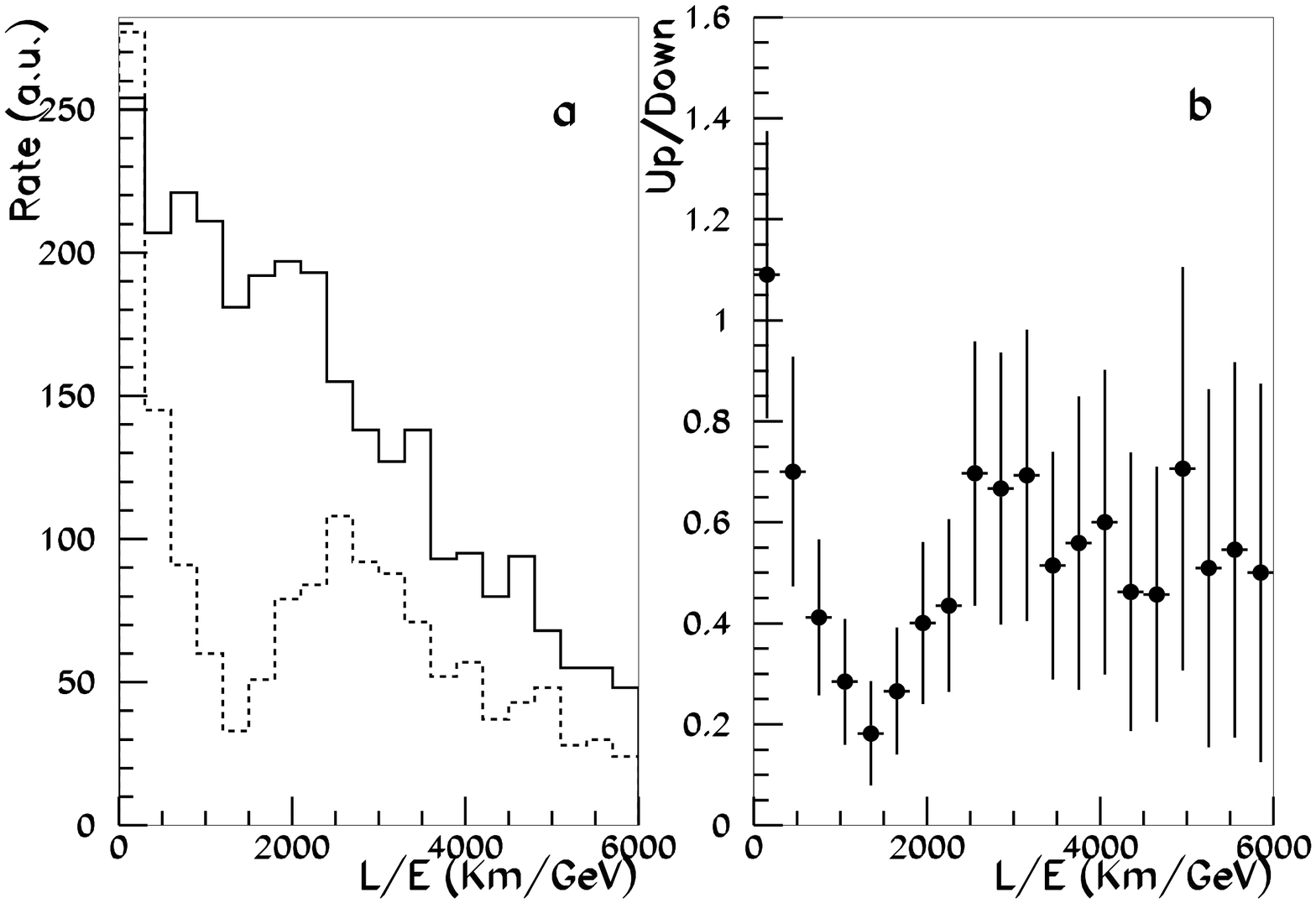,width=\linewidth}}
\vspace{-1.0cm}
\end{center}
\caption{As fig. \ref{fig:one}, $\Delta m^2 = 10^{-3}\ eV^2$}
\label{fig:two}
\end{figure}

\begin{figure}[tb]
\begin{center}
\mbox{\epsfig{file=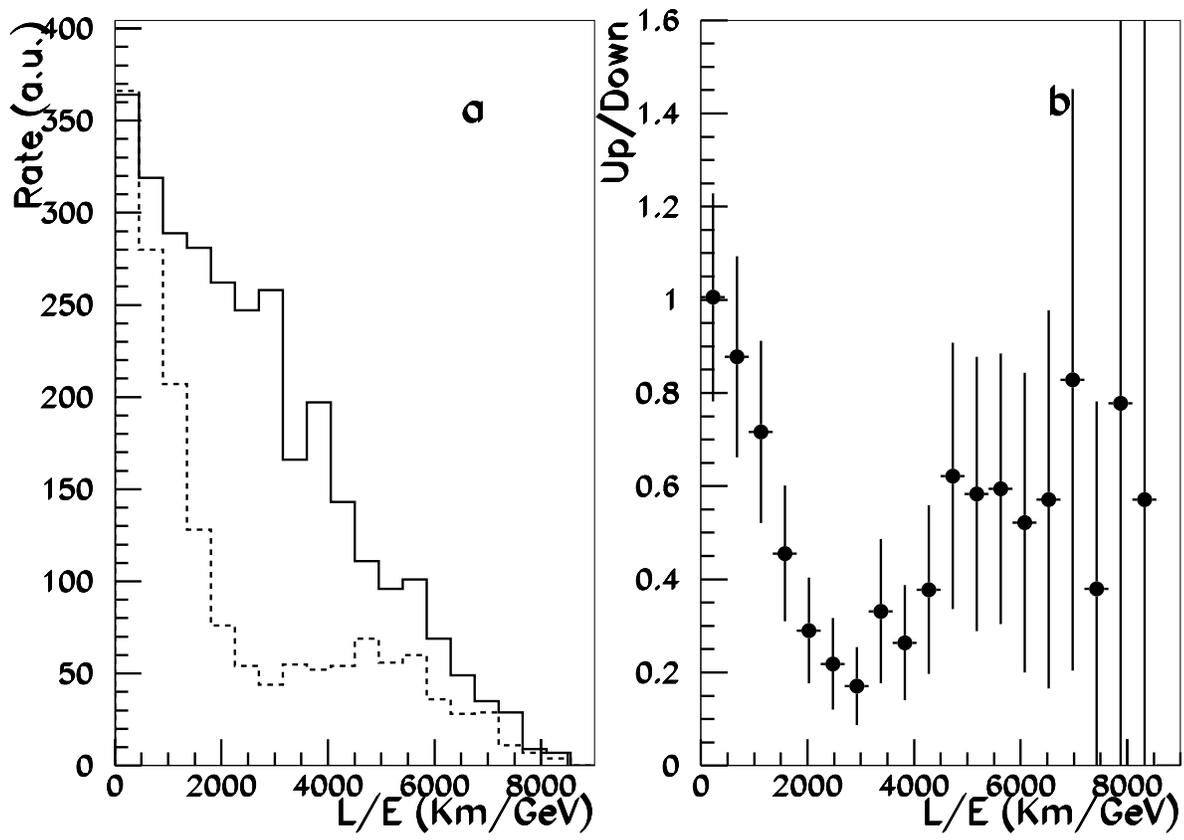,width=\linewidth}}
\end{center}
\caption{As fig. \ref{fig:one}, $\Delta m^2 = 5 \times 10^{-4}\ eV^2$}
\label{fig:three}
\end{figure}

\begin{figure}[tb]
\begin{center}
\mbox{\epsfig{file=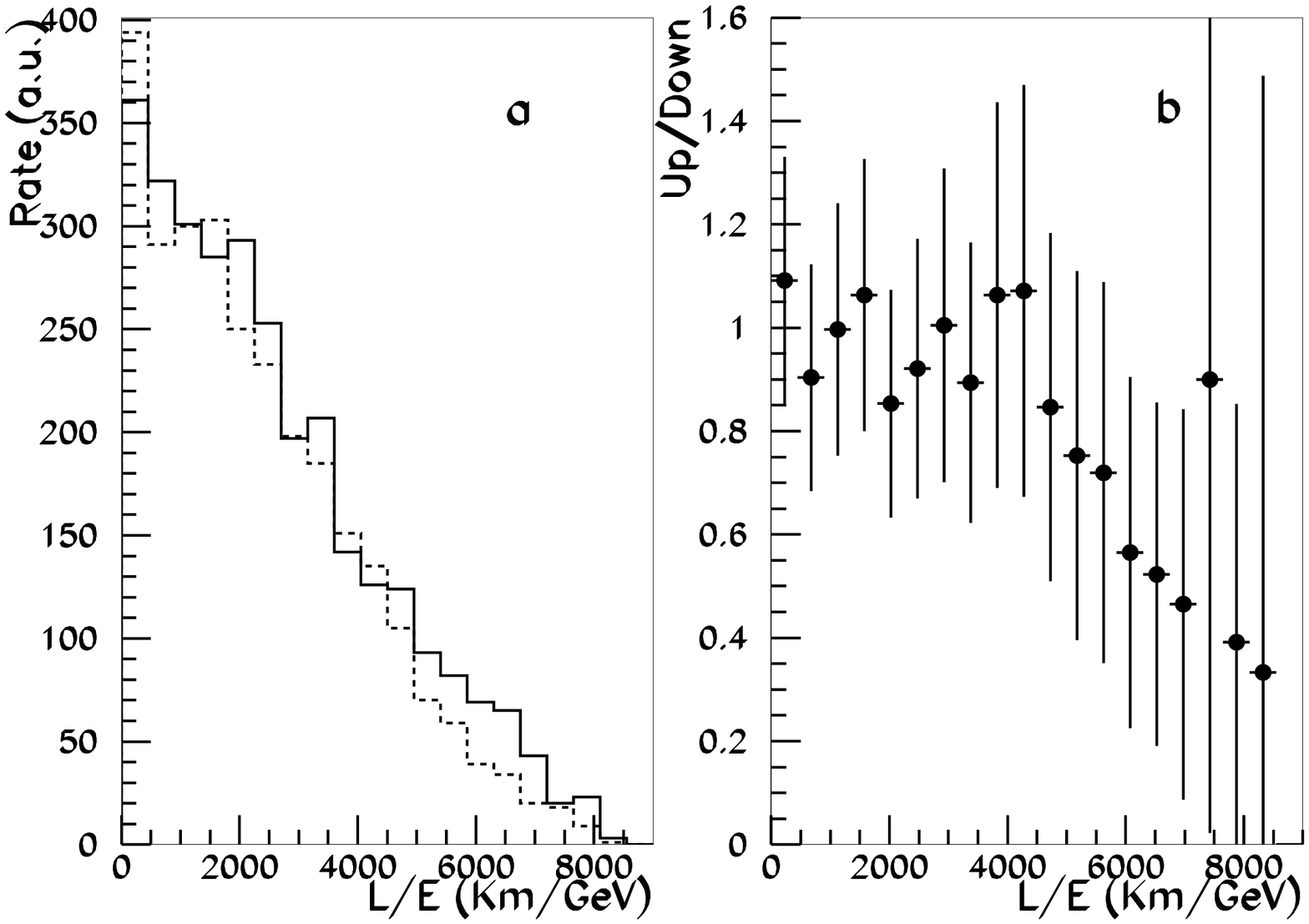,width=\linewidth}}
\end{center}
\caption{As fig. \ref{fig:one}, $\Delta m^2 = 10^{-4}\ eV^2$}
\label{fig:four}
\end{figure}

We present, as examples, the $L/E$ distributions
obtained with this method for several values of $\Delta m^2$:
$5 \times 10^{-3}\ eV^2$ (fig. \ref{fig:one}),
$10^{-3}\ eV^2$ (fig. \ref{fig:two}),
$5 \times 10^{-4}\ eV^2$ (fig. \ref{fig:three}),
$10^{-4}\ eV^2$ (fig. \ref{fig:four}). 
We recall that in order to compare the upward going neutrino sample 
with the downward going one as a function of $L/E$, we have assigned
to the downward going neutrinos (zenith angle $\theta < \pi / 2$) 
the distance they would have traveled if $\theta = \pi - \theta$. 
The ratio of the two distributions is thus 
approximately flux independent and the oscillating pattern is then 
put in evidence.

For $\Delta m^2$ larger than
$5\times 10^{-3}$, the precision on $L/E$ is no more sufficient
to resolve the narrow oscillation pattern, thus the value of
$\Delta m^2$ cannot be measured.
Still the oscillations can be
identified by a deficit of upward with respect to downward events,
which will result in an average ratio of 0.5 in the case of maximal
mixing.

\subsection{Appearance of high-energy muon-less events}

Because of the large deficit of upward with respect to downward muon events,
for $\Delta m^2 > 10^{-3}$ there is a simple method to measure
the $\nu_\tau$
appearance and/or distinguish between  $\nu_\mu\rightarrow \nu_\tau$
oscillation and 
$\nu_\mu\rightarrow \nu_{sterile}$ oscillation
(we assume that a sterile neutrino 
interacts neither via charged currents nor via neutral currents).

The method consists in measuring the up/down ratio of the high energy 
muon-less events, with the vertex clearly identified, as a function of the 
visible energy. 

An event is considered to be muon-less if it does not contain 
non-interacting tracks longer than $1\ m$ (equivalent to $0.9\ GeV$
for a m.i.p.); 
the visible energy is defined as the quadratic sum of the digital hits 
in two orthogonal views; the up/down direction is determined by the shape 
of the hadronic shower development.

\begin{figure}[tb]
\begin{center}
\mbox{\epsfig{file=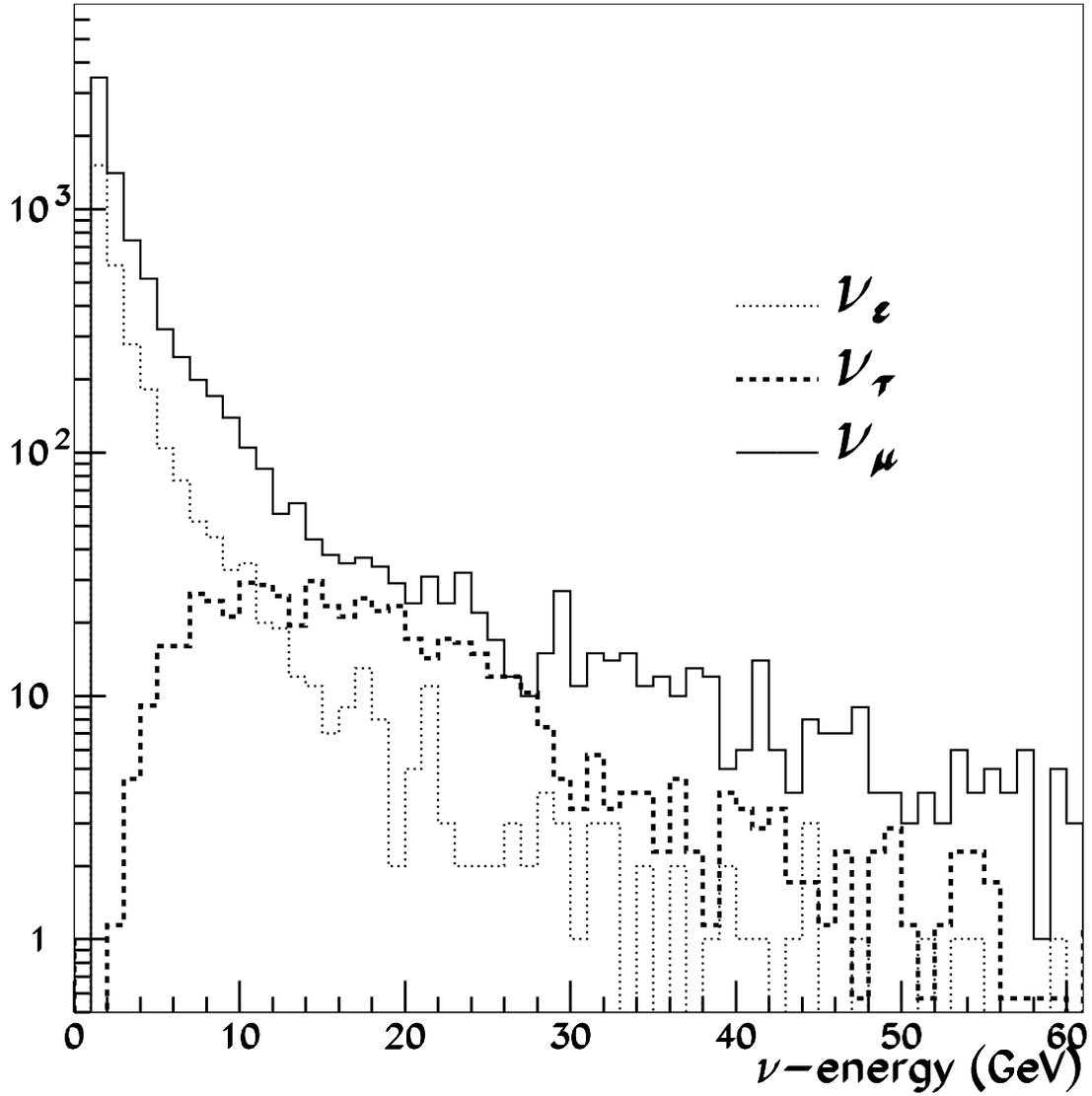,width=\linewidth}}
\end{center}
\caption{Energy spectra for CC interactions produced by neutrinos
of different flavours integrated over the full solid angle. The spectrum
of $\nu_\tau$ events is obtained in the hypothesis of full $\nu_\mu$ into
$\nu_\tau$ conversion.}
\label{nuspectra}
\end{figure}

The energy spectra of the $\nu_\mu$ and $\nu_e$ CC events integrated
over the full
solid angle are shown in fig.~\ref{nuspectra}. This spectra have been provided
by Lipari et al.~\cite{Lipari} as a function of $E$ and $\cos(\theta)$,
where $\theta$ is the zenith angle, and
have been extensively
used in our simulation. The $\nu_\tau$ CC interaction spectrum is also shown 
in fig.~\ref{nuspectra} in the hypothesis of full $\nu_\mu$ conversion.

For sake of clarity, in table~\ref{tab:nuflu1} we give the integrated values
of the neutrino CC events rates for a detector exposure 
of $30\ kt \cdot 4\ y$ and $0 < \cos(\theta) < 1$; in table~\ref{tab:nuflu2}
we give the same rates but for $0.5 < \cos(\theta) < 1$.

\begin{table}[h]
\begin{center}
\begin{tabular}{cccc}
\hline
$E_{min} (GeV)$ & $\nu_\mu$ & $\nu_e$ & $\nu_\tau$ \\
\hline
$1.$ & 3767 & 1654 & 266 \\
$3.$ & 1470 & 505 & 266 \\
$10.$ & 453 & 107 & 179 \\
$30.$ & 121 & 20 & 71 \\
\hline
\end{tabular}
\caption{Neutrino + anti-neutrino integrated CC event rate 
for a detector exposure 
of $30\ kt \cdot 4\ y$ and $0 < \cos(\theta) < 1$; full $\nu_\mu$ 
conversion is assumed for the $\nu_\tau$ case. The integrated NC event 
rate is about one third of the $\nu_\mu$ + $\nu_e$ CC event rate.}
\label{tab:nuflu1}
\end{center}
\end{table}

\begin{table}[h]
\begin{center}
\begin{tabular}{cccc}
\hline
$E_{min} (GeV)$ & $\nu_\mu$ & $\nu_e$ & $\nu_\tau$ \\
\hline
$1.$ & 1677 & 612 & 111 \\
$3.$ & 623 & 157 & 111 \\
$10.$ & 187 & 27 & 74 \\
$30.$ & 49 & 4 & 29 \\
\hline
\end{tabular}
\caption{Same as table~\ref{tab:nuflu1} but for $0.5 < \cos(\theta) < 1$.}
\label{tab:nuflu2}
\end{center}
\end{table}

Given a specific set of oscillation parameters, 
the unoscillated $\nu_\mu$ event 
distribution is obtained simply multiplying the spectrum in
fig.~\ref{nuspectra} by $(1-P)$, where $P$ is given by equation (\ref{prob}), 
the $\nu_\tau$ events distribution is obtained multiplying the
$\nu_\tau$ spectrum of 
fig.~\ref{nuspectra} by $P$. The $\nu_e$ spectrum is assumed to be unaffected 
by oscillations.

From these spectra it is clear that, in order to enhance the $\nu_\tau$
contribution 
to muon-less events, one has to select candidates with high visible energy. 

The $\nu_\mu$ CC events rejection is good at high energy
because of the cut of muon with energy larger that 0.9 GeV (due to the 
flat y distribution of the CC interaction).

The $\nu_e$ CC events rejection is due to the characteristic feature
of the detector to filter off the electro-magnetic component of the 
interaction. As a consequence the visible energy is only due to the 
residual hadronic component as in the case 
of neutral current events.

A very important feature of this detector that helps in reducing the
uncertainty on the up/down muon-less ratio is the non isotropy of the detector.
The reason is the following:
the horizontal events are of little use in the up/down ratio, in fact
they do not oscillate enough, their hemisphere is uncertain and the $\nu_e$
background
is larger. The fact that the detector has a vertical development and a
thick sampling causes the rejection of most of horizontal NC events
and $\nu_e$ CC, because their limited vertical development prevents the
identification of their direction.

We have performed extensive simulations of the detector performance with 
various thicknesses of the absorber layers, from 2 cm up to 10 cm. The 
simulated sample corresponds to a detector exposure of $30\ kt \cdot 4\ y$.
The analysis of the simulated data has been performed first through visual 
scanning to optimise the selection cuts, then in an automated way where the 
best selection cuts have been implemented. 

\begin{figure}[tb]
\begin{center}
\mbox{\epsfig{file=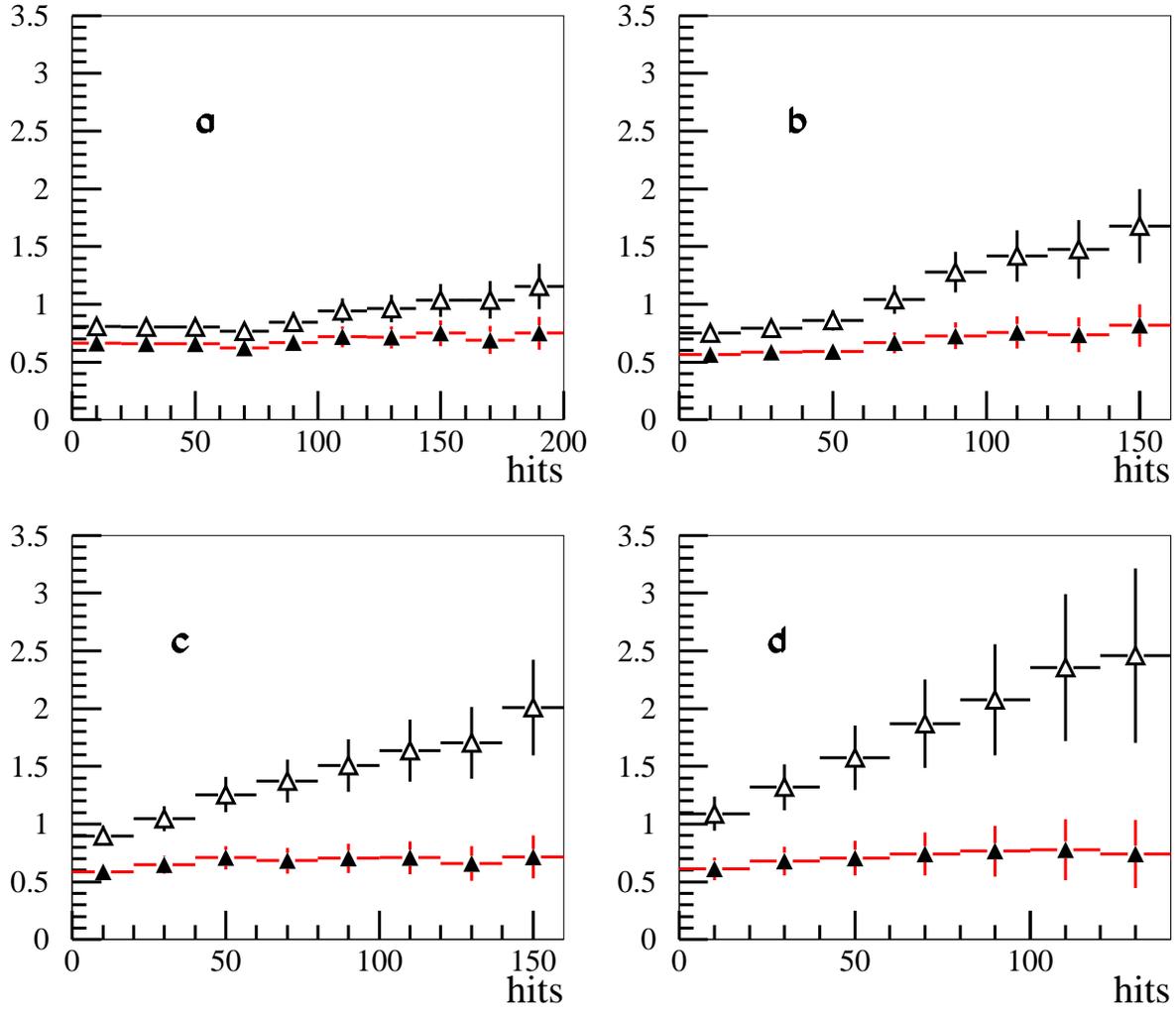,width=\linewidth}}
\end{center}
\caption{Up/down integral ratios for muon-less events,
selected as explained in text,
for $\nu_\mu$ oscillation to a sterile neutrino (full triangles) and to
tau neutrino (open triangles), for $\Delta m^2 = 5\times 10^{-3}\rm\ eV^2 $
and various thicknesses of the absorber layers:
{\it fig.~a 2 cm, fig.~b 4 cm, fig.~c 6 cm, fig.~d 8 cm.}
Error bars correspond to 4 years of data taking.}
\label{updown1}
\end{figure}

We analysed the ratio of up to down events which produce a number of hits
above a given value and satisfy the above criteria.
In fig.~\ref{updown1} we show these integral ratios as a function
of the lower bound on hits for a value of 
$\Delta m^2$ equal to $5\times 10^{-3}\ eV^2$
and several thicknesses of the 
iron absorber.
Both the cases of $\nu_\mu\rightarrow \nu_\tau$ oscillation and
$\nu_\mu\rightarrow\nu_{sterile}$ oscillation 
are shown. The same results are obtained for larger values of $\Delta m^2$.

The choice for the best iron thickness is determined by the separation between 
the two oscillation hypothesis. From a visual inspection of the various plots 
it turns out that any thickness between $4\ cm$ and $8\ cm$ is acceptable. 
In fact, while for small thicknesses the tracking ability is slightly 
enhanced, the electro-magnetic filtering capability is strongly
reduced mainly in the high energy region (where the $\nu_\tau$ events occur). 
It follows that for our purpose a layer thickness of several radiation length 
is preferred. We believe that $8\ cm$ is the best solution because it allows 
to minimise the surface of the active elements (i.e. streamer tubes).

\begin{figure}[tb]
\begin{center}
\mbox{\epsfig{file=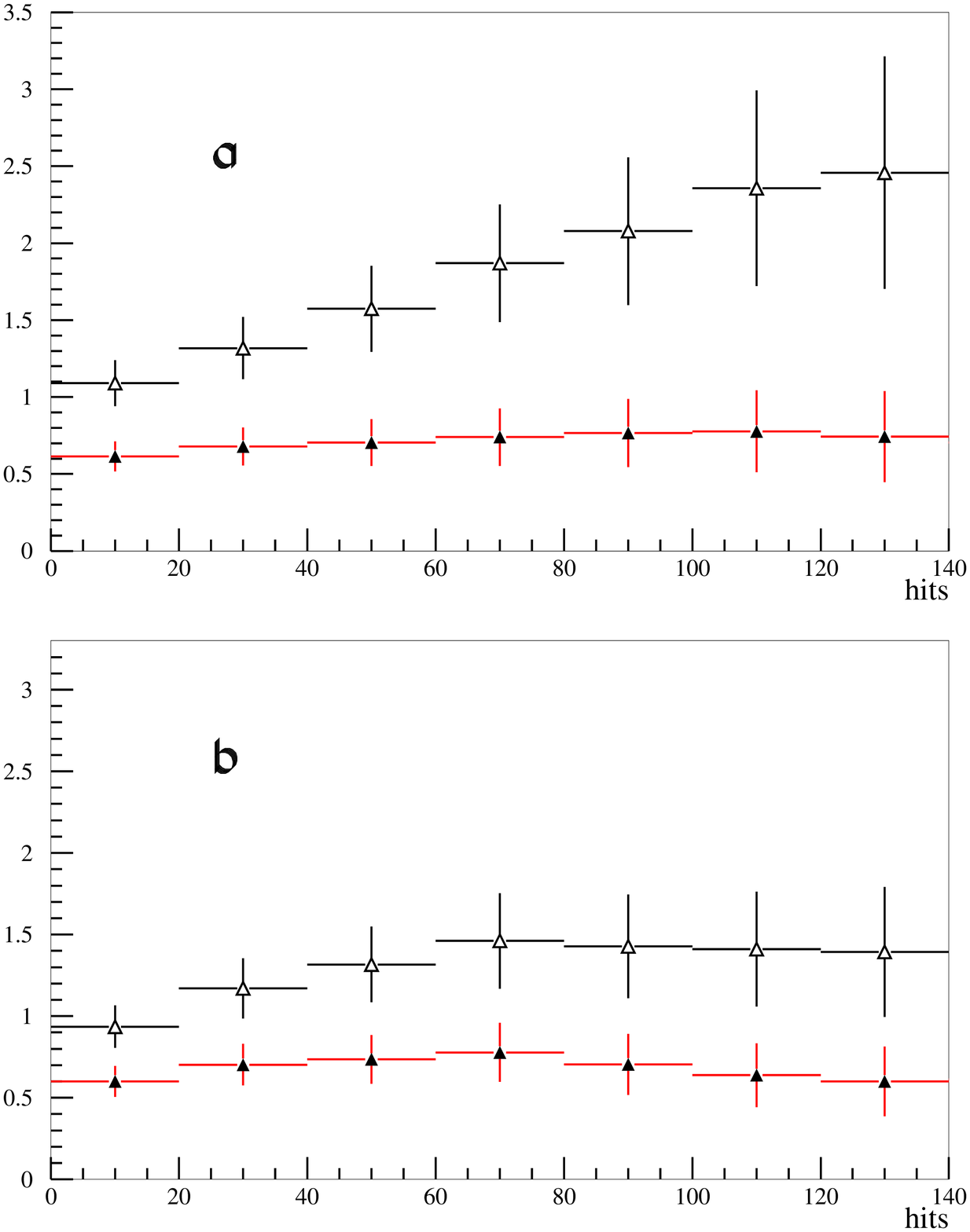,width=15.cm}}
\end{center}
\caption{Up/down integral ratios for muon-less events, 
for $\nu_\mu$ oscillation to a sterile neutrino (full triangles) and to
tau neutrino (open triangles), for a thickness of the iron 
absorber of 8 cm. {\it Fig.~a:} $\Delta m^2 =5\times 10^{-3}\rm\ eV^2 $; 
{\it fig.~b:} $\Delta m^2 =3\times 10^{-3}\rm\ eV^2 $.
Events have been generated with high statistics,
error bars correspond to 4 years of data taking.}
\label{updown2}
\end{figure}

In fig.~\ref{updown2} we show again the up/down integral ratio defined
above for a thickness of the iron absorber of $8\ cm$ and for two values 
of $\Delta m^2$. The statistical separation between the two oscillation cases 
amounts to several standard deviations for $\Delta m^2 > 3\times 10^{-3}$. 
In the $\nu_\mu\rightarrow\nu_\tau$ case there is an excess of muon-less 
events with high visible energy from the bottom hemisphere due to the
semileptonic decay of the tau lepton ($BR \simeq 0.7$) that produce neutral
current like events; in the $\nu_\mu\rightarrow\nu_{sterile}$ case there is a 
lack of neutral currents from the bottom hemisphere at all visible energies 
because the sterile neutrino does not interact via neutral currents. 


\section{Conclusions}
We have demonstrated that a high density detector of 30 kt with rough 
sampling and good tracking
capability is well suited to solve the atmospheric neutrino puzzle as stated 
by the Super-Kamiokande experiment.

This is performed both with  appearance and disappearance methods; the 
oscillation parameters are measured over a wide range of   $\Delta m^2$
values and for the mixing parameter larger that 0.6.


The detector is easily built in a short delay because it does not require
R\&D phases. For the construction one could use the same technique of the
MACRO detector, its infrastructures and active elements could be recycled.

This experiment, completely devoted to the atmospheric neutrino study, is 
complementary to those designed for long base line neutrino detection
\cite{icarusprop,opera,aquarich}.

Its cost is limited (a rough but safe estimate is around 30 GLit half
of which is for the iron). It occupies half of the area taken by
MACRO at present.

We believe that this experiment together with the LBL beam
will put Italy in fore-line in the fascinating quest of the neutrinos'
nature. 

\section*{Acknowledgements}
We gratefully acknowledge A.~Rubbia for providing us the
neutrino event generator, without which this work would not
have been possible. We are also in debt with T.~Ypsilantis for
the valuable discussions that led to the clarification of the ideas 
described in this paper.



\end{document}